\documentclass[aps,prd,onecolumn,amsmath,11pt,superscriptaddress,floatfix,nofootinbib,preprintnumbers]{revtex4-2}

\usepackage{mathrsfs}
\usepackage{amsfonts}
\usepackage{amsmath,amssymb,bm}
\usepackage{array}
\usepackage{verbatim}
\usepackage{epsfig}
\usepackage{graphicx}
\usepackage{hyperref}
\hypersetup{colorlinks, linkcolor = [rgb]{0, 0, 0.5}, citecolor = [rgb]{0,0.0,0.5}, urlcolor = [rgb]{0,0.0,0.5}}
\usepackage[normalem]{ulem}
\usepackage{xcolor}
\usepackage{ulem}
\usepackage{multirow}
\usepackage{mathtools}
\DeclarePairedDelimiter\abs{\lvert}{\rvert}
\usepackage{booktabs}
\usepackage{subcaption}
\usepackage{graphicx}
\graphicspath{{./figs/}}
\usepackage{dcolumn}
\usepackage{bm}
\usepackage{slashed}
\usepackage{siunitx}
\usepackage{ulem,xpatch}
\usepackage{hyperref}
\usepackage[mathlines]{lineno}
\usepackage{comment}
\usepackage{soul}
\usepackage{xcolor}
\usepackage{xfrac}

\allowdisplaybreaks[4]

\begin{document}

\title{Model-Independent and Data-Driven Extraction of the Photon Distribution Function at the Electron--Ion Collider}

\author{Cong Li}
\affiliation{School of Information Engineering, Zhejiang Ocean University, Zhoushan, Zhejiang, China}

\date{\today}


\begin{abstract}
In this work, we investigate a cross-section-ratio-based inversion method for extracting distribution functions at the Electron--Ion Collider (EIC). Starting from the general factorization formula $d\sigma = H \otimes f$, we show that, for an appropriate bremsstrahlung process and suitably chosen differential observables, the convolution structure can be reduced to a multiplicative form. On this basis, we define the ratio $R = \frac{d\sigma_{\mathrm{exp}}}{d\sigma_{\mathrm{hard}}},$ where \(d\sigma_{\mathrm{exp}}\) is the experimentally measured differential cross section and \(d\sigma_{\mathrm{hard}}\) is a perturbatively calculable theoretical input. We then establish the relation between \(R\) and the target distribution function \(f\). Finally, we investigate how the theoretical input \(d\sigma_{\mathrm{hard}}\) should be constructed when soft-photon radiation, finite-bin-width effects, and experimental acceptance are taken into account.

\end{abstract}

\maketitle

\section{Introduction}
\label{sec:introduction}

Nucleon structure distributions encode the nonperturbative dynamics of the strong interaction~\cite{Gao:2018}. In high-energy scattering, experimentally measurable differential cross sections can often be factorized into a short-distance hard-scattering contribution and a long-distance structure contribution~\cite{Collins:1985}. The former can be calculated systematically in perturbative QCD or QED, whereas the latter is absorbed into the corresponding distribution functions. Reliable extraction of these nonperturbative distributions from measured cross sections is therefore a central problem in hadron and nuclear structure physics~\cite{Gao:2018}. From a theoretical perspective, such an extraction is intrinsically an inverse problem~\cite{DelDebbio:2022}. A generic factorized cross section may be written schematically as
\begin{equation}
    d\sigma = H \otimes f,
\end{equation}
where \(H\) denotes the perturbatively calculable hard-scattering kernel, \(f\) is the nonperturbative distribution to be determined, and \(\otimes\) represents a convolution over the relevant kinematic variables. Conventional global analyses adopt a forward-fitting strategy: a parametrized form is assumed for \(f\), inserted into the factorization formula, and constrained by comparison with experimental data~\cite{Gao:2018,Ball:2022}. This framework has been highly successful in analyses of collinear PDFs~\cite{Ball:2022}, nuclear PDFs~\cite{Eskola:2022}, transverse-momentum-dependent distributions~\cite{MAP:2022}, and spin-dependent distributions~\cite{deFlorian:2009}, and remains the standard approach in nucleon-structure phenomenology.

Nevertheless, forward fitting inevitably introduces model and parametrization dependence~\cite{DelDebbio:2022,Carrazza2022}. Since the functional form of a nonperturbative distribution is not fixed by perturbation theory, assumptions must be made about its dependence on \(x\), \(k_\perp\), and other kinematic variables. This creates two generic difficulties. First, complicated background fields and nonperturbative structures may be difficult to represent with a predetermined functional form. Second, the number of fit parameters must balance flexibility against statistical stability~\cite{DelDebbio:2022}: too few parameters may bias the extraction, while too many require substantially larger data samples and may lead to poorly constrained fits~\cite{Ball:2015}. These issues are particularly relevant for nuclear and transverse-momentum-dependent distributions~\cite{Khalek2019,Scimemi2020}, where the true structure may contain non-Gaussian tails, nontrivial \(x\)-\(k_\perp\) correlations~\cite{Scimemi2020,Bury2022}, nuclear-medium modifications, or localized features. Strong parametrization priors may smooth out or obscure such effects~\cite{Carrazza2022,Bury2022}. It is therefore desirable to develop a more direct, data-driven method that minimizes assumptions about the functional form of the target distribution and reconstructs nonperturbative information directly from measured observables. 

In this work, we investigate a cross-section-ratio method for inverting distribution functions in electron--hadron collisions. Instead of following the conventional sequence
\begin{equation}
    \text{parametrize } f
    \;\longrightarrow\;
    \text{compute } H\otimes f
    \;\longrightarrow\;
    \text{fit the data},\notag
\end{equation}
we begin with a measured final-state differential cross section and divide it by a perturbatively calculable hard contribution. We define
\begin{equation}
    R =
    \frac{d\sigma_{\mathrm{exp}}}
         {d\sigma_{\mathrm{hard}}},
\end{equation}
where \(d\sigma_{\mathrm{hard}}\) contains the hard-scattering kernel and the relevant kinematic factors. For suitably chosen bremsstrahlung processes, the nonperturbative background-field dependence and the hard-scattering contribution can enter the differential cross section multiplicatively at fixed final-state rapidity and transverse momentum. In the corresponding QCD process, a quark pair traversing a gluonic background field may radiate a gluon that carries information about that field. The ratio \(R\) then removes the hard factor and isolates the target gluon distribution~\cite{PhysRevD.83.105005,PhysRevLett.132.081902}.

The present study focuses on the mathematical consistency of this inversion method, its stability in an initial phenomenological implementation, and the construction of the theoretical input \(d\sigma_{\mathrm{hard}}\). Rather than studying gluon radiation from a quark pair in a gluonic background field directly, we first consider the QED analogue: photon radiation from an electron--positron pair propagating through an electromagnetic background field \cite{PhysRevD.101.075049}. This choice is motivated by three considerations. First, QED provides a close Abelian analogue of the corresponding QCD dynamics, with the quark pair replaced by a lepton pair and the gluonic background field replaced by a photon field. Second, the QED process is theoretically cleaner, while electrons, positrons, and photons can be measured with high precision. Third, the electromagnetic field surrounding a relativistic nucleus can be described accurately within the equivalent-photon approximation~\cite{BUDNEV1975181,BERTULANI1988299,Jentschura2009,Li1,Li2}, providing a controlled reference against which the extracted photon distribution can be tested.

We propose to implement this method at the future Electron--Ion Collider, whose principal goals include precision studies of nucleon and nuclear structure~\cite{ABDU,Accardi2016}. The process of interest is initiated by the production of a timelike virtual photon through a Bethe--Heitler-type subprocess. The virtual photon subsequently decays into an electron--positron pair, and the pair radiates a photon while propagating through the electromagnetic field of the nucleus. The differential cross section for the complete process corresponds to \(d\sigma_{\mathrm{exp}}\). By exploiting its multiplicative structure and dividing out the perturbative hard contribution, one can extract the photon distribution associated with the nuclear electromagnetic field. We then extend the construction to more realistic situations. In particular, we study how \(d\sigma_{\mathrm{hard}}\) should be defined when soft-photon radiation and Sudakov effects are included \cite{YENNIE1961379,PhysRevD.102.094013,PhysRevD.108.116015}, and how the extraction is modified by finite bin widths, kinematic cuts, detector acceptance, and other experimental effects.

The remainder of this paper is organized as follows. In Sec.~\ref{sec2}, we derive the Bethe--Heitler-induced lepton-pair production cross section and define the corresponding virtual-photon distribution, together with its convolution with the subsequent subprocess. In Sec.~\ref{sec3}, we study the convolution of the Bethe--Heitler virtual-photon distribution with $\gamma^* \to e^+e^-\gamma$ and demonstrate the mathematical validity of extracting the background-field photon distribution through the ratio \(R\). In Sec.~\ref{sec4}, we discuss the construction of \(d\sigma_{\mathrm{hard}}\) in the presence of Sudakov effects. In Sec.~\ref{sec5}, we incorporate realistic experimental effects into the hard input. Section~\ref{sec6} summarizes our conclusions.

\section{Bethe--Heitler-Induced Lepton-Pair Production and the Virtual-Photon Distribution}
\label{sec2}
In this section, we discuss the distribution of timelike virtual photons generated by a Bethe--Heitler-type subprocess at the Electron--Ion Collider. We also show that, when the azimuthal distribution of the final-state lepton pair is not measured, the production and decay tensors can be reduced to a scalar factorized form.

We consider the incoherent equivalent-photon approximation (EPA) for the process
\begin{equation}
    e(l)+A(P_A)
    \to
    e(l')+e^+(p_1)+e^-(p_2)+A(P_A').
\end{equation}
Within the EPA, the nucleus supplies a quasi-real photon,
\(\gamma_A(k)\), and the corresponding hard subprocess is
\begin{equation}
    e(l)+\gamma_A(k)
    \to
    e(l')+e^+(p_1)+e^-(p_2).
\end{equation}
It is important to distinguish the process of our interest,
\begin{align}
    e(l)+\gamma_A(k)
    &\to e(l')+\gamma^*(q),
    \\
    \gamma^*(q)
    &\to e^+(p_1)+e^-(p_2),
\end{align}
from the crossed, or ``twin,'' mechanism
\begin{align}
    e(l)
    &\to e(l')+\gamma^*(q),
    \\
    \gamma^*(q)+\gamma_A(k)
    &\to e^+(p_1)+e^-(p_2).
\end{align}
To suppress the nearly collinear photon emission from the electron line in the latter mechanism, we impose a lower cut on the electron-line virtuality,
\begin{equation}
    Q_e^2
    \equiv
    -(l-l')^2
    >
    Q_{e,\min}^2.
\end{equation}
This cut removes the electron-EPA-enhanced region \(Q_e^2\to 0\). Increasing \(Q_{e,\min}^2\) leads to stronger suppression of the competing mechanism, at the cost of a reduced event yield. We also require the final-state electron to carry an observable transverse recoil,
\begin{equation}
    \abs{\bm l'_\perp}>l_{\perp,\min},
\end{equation}
thereby selecting the Bethe--Heitler hard-scattering topology $e+\gamma_A\to e'+\gamma^*.$ A larger value of \(l_{\perp,\min}\) further suppresses the competing topology, but also removes more signal events. The restriction to a Bethe--Heitler-dominated kinematic region is introduced here primarily to present the subsequent extraction method in a transparent and mathematically controlled setting. Residual contributions from the competing subprocess act as a systematic uncertainty. In a precision analysis, this uncertainty can be eliminated by including the competing subprocess explicitly in the future.

\subsection{Tensor distribution of Bethe--Heitler virtual photons}

We first consider the production subprocess and define the amplitude with open Lorentz indices as
\begin{equation}
    \mathcal{M}_{\mathrm{BH}}^{\mu\alpha}
    \equiv
    \mathcal{M}^{\mu\alpha}
    \bigl(
        e(l)+\gamma_A(k)
        \to
        e(l')+\gamma^*(q)
    \bigr),
\end{equation}
where \(\alpha\) is the Lorentz index of the incoming EPA photon and \(\mu\) is that of the outgoing timelike virtual photon. The corresponding Bethe--Heitler hard tensor is defined by
\begin{equation}
    H_{\mathrm{BH}}^{\mu\nu;\alpha\beta}(l,k,q)
    =
    \frac{1}{2}
    \sum_{\mathrm{spins}}
    \mathcal{M}_{\mathrm{BH}}^{\mu\alpha}
    \left(
        \mathcal{M}_{\mathrm{BH}}^{\nu\beta}
    \right)^*,
\end{equation}
where the factor \(1/2\) denotes the average over the initial-state electron spin.

The polarization and transverse-momentum structure of the incoming EPA photon are encoded in the nuclear photon correlator
\(\Phi_{\gamma/A}^{\alpha\beta}\). Retaining only the unpolarized photon distribution, one may write
\begin{equation}
    \Phi_{\gamma/A}^{\alpha\beta}
    (x_\gamma,\bm k_\perp)
    \longrightarrow
    -\frac{1}{2}
    g_\perp^{\alpha\beta}\,
    x_\gamma
    f_1^{\gamma/A}
    (x_\gamma,k_\perp^2).
\end{equation}
We then define the tensor distribution of virtual photons produced through the Bethe--Heitler subprocess as
\begin{align}
    d\mathcal{F}_{\mathrm{BH}}^{\mu\nu}(q)
    &=
    \int
    dx_\gamma\,d^2\bm k_\perp\,
    \Phi_{\gamma/A}^{\alpha\beta}
    (x_\gamma,\bm k_\perp)
    \frac{1}{2\hat{s}}\,
    H_{\mathrm{BH}}^{\mu\nu;\alpha\beta}(l,k,q)
    d\Omega_2(l+k;l',q),
    \label{eq:BH-tensor-distribution}
\end{align}
where
\begin{align}
    d\Omega_2(l+k;l',q)
    &=
    \frac{d^3\bm l'}{(2\pi)^3\,2E_{l'}}
    \frac{d^3\bm q}{(2\pi)^3\,2E_q}
    (2\pi)^4
    \delta^{(4)}(l+k-l'-q).\notag
\end{align}
The quantity \(\mathcal{F}_{\mathrm{BH}}^{\mu\nu}\) is the source tensor for virtual photons generated in the production subprocess. Since it carries two Lorentz indices, it is more precisely a polarization density matrix rather than an ordinary scalar distribution. The produced virtual photon therefore retains its full polarization information. Electromagnetic current conservation implies the transversality conditions
\begin{equation}
    q_\mu\,d\mathcal{F}_{\mathrm{BH}}^{\mu\nu}(q)=q_\nu\,d\mathcal{F}_{\mathrm{BH}}^{\mu\nu}(q)=0.
\end{equation}
For a timelike virtual photon with \(q^2=Q^2>0\), the physical polarization sum is
\begin{equation}
    P_{\mathrm{phys}}^{\mu\nu}(q)
    =
    \sum_{\lambda=0,\pm1}
    \epsilon_\lambda^\mu(q)
    \epsilon_\lambda^{*\nu}(q)
    =
    -g^{\mu\nu}
    +
    \frac{q^\mu q^\nu}{Q^2}.
\end{equation}
Since a timelike virtual photon has three physical polarization states,
\begin{equation}
    P_{\mathrm{phys}}^{\mu\nu}(q)
    P_{\mathrm{phys},\mu\nu}(q)
    =
    3.
    \label{17}
\end{equation}
When only the unpolarized cross section is considered and neither the virtual-photon polarization nor the angular distribution of the final-state lepton pair is measured, the tensor source can be projected onto a scalar virtual-photon distribution:
\begin{equation}
    d\mathcal{F}_{\mathrm{BH}}^{\gamma^*}(q)
    =
    P_{\mathrm{phys},\mu\nu}(q)\,
    d\mathcal{F}_{\mathrm{BH}}^{\mu\nu}(q).
    \label{eq:scalar-BH-distribution}
\end{equation}
Using transversality, this becomes
\begin{equation}
    d\mathcal{F}_{\mathrm{BH}}^{\gamma^*}(q)
    =
    -g_{\mu\nu}\,
    d\mathcal{F}_{\mathrm{BH}}^{\mu\nu}(q).
\end{equation}

\subsection{Decay tensor and scalar reduction}

We next consider the decay subprocess
\begin{equation}
    \gamma^*(q)\to e^+(p_1)+e^-(p_2).
\end{equation}
The corresponding leptonic tensor is denoted by $L_{\mu\nu}(p_1,p_2).$ Including the virtual-photon propagator, the decay vertex, and the two-body phase space, we define the decay kernel
\begin{equation}
    d\mathcal{D}_{\mu\nu}(q)
    =
    \frac{dQ^2}{2\pi}
    \frac{e^2}{Q^4}
    L_{\mu\nu}(p_1,p_2)\,
    d\Omega_2(q;p_1,p_2).
\end{equation}
The most general factorized expression for the full process is therefore
\begin{equation}
    d\sigma
    \bigl(
        eA\to e\,e^+e^-A
    \bigr)
    =
    d\mathcal{F}_{\mathrm{BH}}^{\mu\nu}(q)\,
    d\mathcal{D}_{\mu\nu}(q).
    \label{eq:tensor-convolution}
\end{equation}
Equation~\eqref{eq:tensor-convolution} is a tensor contraction. The production subprocess provides the virtual-photon polarization density matrix
\(d\mathcal{F}_{\mathrm{BH}}^{\mu\nu}\), while the decay subprocess provides the tensor
\(d\mathcal{D}_{\mu\nu}\). The two are coupled through the Lorentz indices of the intermediate virtual photon.

If the decay angles of the \(e^+e^-\) pair are retained,
\(d\mathcal{D}_{\mu\nu}\) is not isotropic and cannot be reduced to a scalar factor before contraction. By contrast, if the azimuthal distribution is not measured and the decay angles are fully integrated, there is no preferred spatial direction in the virtual-photon rest frame. The angle-integrated leptonic tensor must then be proportional to the physical polarization projector:
\begin{equation}
    \int
    d\Omega_2(q;p_1,p_2)\,
    L_{\mu\nu}(p_1,p_2)
    =
    C_1(Q^2)\,
    P_{\mathrm{phys},\mu\nu}(q).
    \label{eq:leptonic-tensor-decomposition}
\end{equation}
The scalar coefficient is
\begin{equation}
    C_1(Q^2)
    =
    \frac{1}{3}
    \int
    d\Omega_2(q;p_1,p_2)\,
    L_{\mu\nu}(p_1,p_2)\,
    P_{\mathrm{phys}}^{\mu\nu}(q).
    \label{c1}
\end{equation}
where the 1/3 comes from Eq.~\eqref{17}. Before performing the \(Q^2\) integration, the angle-integrated decay kernel can therefore be written as
\begin{equation}
    d\mathcal{D}_{\mu\nu}^{\mathrm{before}}(q)
    =
    C_2(Q^2)\,
    P_{\mathrm{phys},\mu\nu}(q),
\end{equation}
with
\begin{equation}
    C_2(Q^2)
    =
    \frac{1}{2\pi}
    \frac{e^2}{Q^4}
    C_1(Q^2).
\end{equation}
Substituting this expression into Eq.~\eqref{eq:tensor-convolution}, one obtains
\begin{align}
    d\sigma
    &=
    d\mathcal{F}_{\mathrm{BH}}^{\mu\nu}(q)\,
    C_2(Q^2)\,
    P_{\mathrm{phys},\mu\nu}(q)=
    d\mathcal{F}_{\mathrm{BH}}^{\gamma^*}(q)\,
    C_2(Q^2).
\end{align}
Thus, after integration over the lepton decay angles, the tensor contraction reduces to a product of two scalar quantities.

The differential cross section may consequently be written as
\begin{equation}
    \frac{
        d\sigma
        \bigl(
            eA\to e\,e^+e^-A
        \bigr)
    }{
        dQ^2\,
        dy_{\gamma^*}\,
        d^2\bm q_\perp
    }
    =
    C_2(Q^2)\,
    \frac{
        d\mathcal{F}_{\mathrm{BH}}^{\gamma^*}
    }{
        dy_{\gamma^*}\,
        d^2\bm q_\perp
    },
    \label{eq:scalar-factorization}
\end{equation}
where
\begin{align}
    \frac{
        d\mathcal{F}_{\mathrm{BH}}^{\gamma^*}
    }{
        dy_{\gamma^*}\,
        d^2\bm q_\perp
    }
    &=
    P_{\mathrm{phys},\mu\nu}(q)
    \frac{
        d\mathcal{F}_{\mathrm{BH}}^{\mu\nu}
    }{
        dy_{\gamma^*}\,
        d^2\bm q_\perp
    }
   =
    -g_{\mu\nu}
    \frac{
        d\mathcal{F}_{\mathrm{BH}}^{\mu\nu}
    }{
        dy_{\gamma^*}\,
        d^2\bm q_\perp
    }.
\end{align}
and $y_{\gamma^*}$ is the rapidity of final virtual-photon. The Bethe--Heitler cross section has been studied extensively. Using the result of Ref.~\cite{Sun_2020}, one finds
\begin{align}
    \frac{
        d\mathcal{F}_{\mathrm{BH}}^{\gamma^*}
    }{
        dy_{\gamma^*}\,
        d^2\bm P_\perp\,
        d^2\bm q_\perp
    }
    &=
    -g_{\mu\nu}
    \frac{
        d\mathcal{F}_{\mathrm{BH}}^{\mu\nu}
    }{
        dy_{\gamma^*}\,
        d^2\bm P_\perp\,
        d^2\bm q_\perp
    }
    =
    H_{\mathrm{BH}}^{\gamma^*}\,
    x_\gamma
    f_1^{\gamma/A}
    (x_\gamma,k_\perp^2),
    \label{eq:BH-factorized-result}
\end{align}
where
\begin{equation}
    H_{\mathrm{BH}}^{\gamma^*}
    =
    2\alpha_{\mathrm{e}}^2 z^2
    \left[
        \frac{
            1+(1-z)^2
        }{
            \bigl[
                q_\perp^2+(1-z)Q^2
            \bigr]^2
        }
        -
        \frac{
            2Q^2q_\perp^2z^2(1-z)
        }{
            \bigl[
                q_\perp^2+(1-z)Q^2
            \bigr]^4
        }
    \right].
\end{equation}
Here, \(\gamma\) denotes the incoming quasi-real photon and \(\gamma^*\) the outgoing timelike virtual photon. Equation~\eqref{eq:BH-factorized-result} is derived in the correlation limit,
\begin{equation}
    \bm P_\perp
    =
    \bm k_\perp
    =
    \bm l'_\perp+\bm q_\perp
    \simeq 0,
    \qquad
    P_\perp\ll q_\perp.
\end{equation}
In this limit, the transverse momenta of the outgoing electron and virtual photon satisfy
\begin{equation}
    \bm q_\perp
    \simeq
    -\bm l'_\perp
    \simeq
    \frac{
        \bm q_\perp-\bm l'_\perp
    }{2}.
\end{equation}
To ensure that \(\bm q_\perp\) denotes the transverse momentum of the virtual photon, our definitions of \(\bm P_\perp\) and \(\bm q_\perp\) are interchanged relative to those used in Ref.~\cite{Sun_2020}.

In summary, the fully angle-integrated decay tensor is decomposed into the scalar coefficient \(C_2\) and the tensor basis
\(P_{\mathrm{phys}}^{\mu\nu}\). Contracting this tensor basis with the Bethe--Heitler production tensor defines the scalar Bethe--Heitler virtual-photon distribution
\begin{equation}
    f_{\mathrm{BH}}^{\gamma^*}
    (y_{\gamma^*},Q^2,q_\perp^2)
    \equiv
    \frac{
        d\mathcal{F}_{\mathrm{BH}}^{\gamma^*}
    }{
        dy_{\gamma^*}\,
        d^2\bm q_\perp
    }.
\end{equation}
The full cross section then takes the scalar factorized form
\begin{equation}
    \frac{
        d\sigma
    }{
        dQ^2\,
        d^2\bm q_\perp\,
        dy_{\gamma^*}
    }
    =
    f_{\mathrm{BH}}^{\gamma^*}
    (y_{\gamma^*},Q^2,q_\perp^2)\,
    C_2
    (y_{\gamma^*},Q^2,q_\perp^2).
\end{equation}
Equivalently, the hard decay factor is obtained by contracting the decay tensor with the polarization average $\frac{1}{3} P_{\mathrm{phys}}^{\mu\nu}(q)$, as shown in Eq.~\eqref{c1} This construction converts the original tensor contraction into a scalar product whenever the angular information of the final-state lepton pair is integrated out.

\section{Photon Radiation in a Background Field and the Extraction of the Photon Distribution Function}
\label{sec3}
\textbf{In Sec.~\ref{sec2}, we constructed the virtual-photon distribution function
$f_{\mathrm{BH}}^{\gamma^\ast}(y_{\gamma^\ast},Q^2,q_\perp^2)$ by integrating over the momenta $p_1$, $p_2$, $k$, and $l$. In this section, these momentum variables are reintroduced with new physical meanings, as specified in Eq.~\eqref{eq:dipole-bremsstrahlung}.}

At an electron--ion collider (EIC), an electron--positron dipole can undergo bremsstrahlung in the peripheral electromagnetic field of a heavy ion. The corresponding cross section carries information about the background field. Therefore, the photon distribution surrounding the nucleus can be extracted from the measured final-state photon spectrum, providing access to photon states in the small-$x$ region. In this process, a virtual photon emitted by the electron splits into an electron--positron pair, which subsequently radiates a real photon under the influence of the peripheral photon field of the heavy ion. The spectrum of the final-state real photon therefore encodes information about the background field and can be used to extract its photon distribution function. We consider the bremsstrahlung process
\begin{equation}
    \gamma^\ast(q)+A(l)
    \rightarrow
    e^+(p_1)+e^-(p_2)+\gamma(k)+A(l).
    \label{eq:dipole-bremsstrahlung}
\end{equation}
Here, $k$ denotes the four-momentum of the real photon radiated by the lepton pair produced in the virtual-photon decay.\textbf{ The symbol $k$ is reused in this section; in the preceding section, it denoted the initial-state photon in the Bethe--Heitler process and was integrated out.}

In this process, the peripheral electromagnetic field of the heavy ion provides a dense photon background. When an electron propagates through this field, it is decelerated by the background and radiates photons. To describe this effect, we construct the photon propagator in the background field, with details given in Appendix~A. Its on-shell, or cut, form is
\begin{equation}
    D_{\mu\nu}^{(\mathrm{cut};\mathrm{bg})}(k)
    =
    2\pi\theta(k^0)\delta(k^2)
    \left(-g_{\mu\nu}\right)
    \left[1+n(\boldsymbol{k})\right].
    \label{eq:cut-bg-propagator}
\end{equation}
Here, $n(\boldsymbol{k})$ is the momentum-space occupation-number distribution of the background photons. Consequently, the radiation of final-state photons is modified only in the heavy-ion direction and not in the electron direction. We define the heavy-ion direction as the positive-rapidity direction and retain only events in that hemisphere. The effect of the photon background is then incorporated by replacing the vacuum cut photon propagator,
$2\pi\theta(k^0)\delta(k^2)(-g_{\mu\nu})$, with Eq.~\eqref{eq:cut-bg-propagator}.

At leading order, the differential cross section for photon radiation from the lepton pair in the background field is
\begin{align}
&\frac{\mathrm{d}\sigma_{\mathrm{bg}}}
{\mathrm{d}y_1\,\mathrm{d}y_2\,\mathrm{d}y_\gamma\,
 \mathrm{d}^2p_{1\perp}\,\mathrm{d}^2p_{2\perp}\,
 \mathrm{d}^2k_\perp}
\nonumber\\
&\quad =
\frac{e^4}{2^3(2\pi)^9}
\int \mathrm{d}y_{\gamma^\ast}\,\mathrm{d}^2q_\perp\,
\frac{H_{\mathrm{Born}}}{Q^4}\,
f_{\mathrm{BH}}^{\gamma^\ast}
\left(y_{\gamma^\ast},Q^2,q_\perp^2\right)
\theta(-y_{\gamma^\ast})
\left[1+n(\boldsymbol{k})\right]
(2\pi)^4
\delta^{(4)}(q-p_1-p_2-k)
\nonumber\\
&\quad =
\frac{\alpha_e^2}{2(2\pi)^3}
\int \mathrm{d}y_{\gamma^\ast}\,\mathrm{d}^2q_\perp\,
\frac{H_{\mathrm{Born}}}{Q^4}\,
f_{\mathrm{BH}}^{\gamma^\ast}
\left(y_{\gamma^\ast},Q^2,q_\perp^2\right)
\theta(-y_{\gamma^\ast})
\left[1+n(\boldsymbol{k})\right]
\delta^{(4)}(q-p_1-p_2-k)
\nonumber\\
&\quad =
\frac{\alpha_e^2}{2(2\pi)^3}
\int \mathrm{d}y_{\gamma^\ast}\,\mathrm{d}^2q_\perp\,
\frac{H_{\mathrm{Born}}}{Q^4}\,
f_{\mathrm{BH}}^{\gamma^\ast}
\left(y_{\gamma^\ast},Q^2,q_\perp^2\right)
\theta(-y_{\gamma^\ast})
\left[1+n(y_\gamma,|k_\perp|)\right]
\delta^{(4)}(q-p_1-p_2-k).
\label{eq:bg-differential-cross-section}
\end{align}
The Born-level hard factor is
\begin{equation}
    H_{\mathrm{Born}}
    =
    \frac{-g_{\mu\nu}}{3}
    \left(-g_{\alpha\beta}\right)
    M^{\mu\alpha}M^{\ast\,\nu\beta}
    =
    \frac{8Q^2(1-z)\bigl[(z-2)z+2\bigr]}
    {3P_\perp^2z^2}
    -\frac{16}{3}.
    \label{eq:born-hard-factor}
\end{equation}
Here, $M^{\mu\alpha}$ is the scattering amplitude, where $\mu$ and $\alpha$ are the Lorentz indices of the incoming and outgoing photons, respectively. The quantity $Q^2$ is the virtuality of the incoming photon, and $z$ is the longitudinal-momentum fraction of the virtual photon carried by the final-state real photon. The factor $-g_{\mu\nu}/3$ follows from $P_{\mathrm{phys},\mu\nu}(q)/3$ after applying the Ward identity.

In evaluating $H_{\mathrm{Born}}$, we employ the correlation limit, in which the electron and positron are approximately back-to-back in the transverse plane. In this limit, the total transverse momentum of the pair,
\begin{equation}
    p_\perp
    =
    q_\perp-k_\perp
    =
    p_{1\perp}+p_{2\perp},
\end{equation}
is small, so that
\begin{equation}
    p_{1\perp}\simeq -p_{2\perp}\simeq P_\perp=\frac{p_{1\perp} -p_{2\perp}}{2}
\end{equation}
The individual transverse momenta $q_\perp$ and $k_\perp$ are also small, typically of order $0$--$50~\mathrm{MeV}$, and the relevant hierarchy is
\begin{equation}
    \max(p_\perp,q_\perp,k_\perp)\ll P_\perp.
    \label{eq:correlation-hierarchy}
\end{equation}
The factor $\theta(-y_{\gamma^\ast})$ forces the virtual photon to propagate in the electron direction. This condition prevents the initial virtual-photon emission from being modified by the background field and thereby avoids introducing additional background-field effects into the production subprocess.

By integrating over the three-momenta of the final-state electron and positron, we obtain the final-state photon spectrum:
\begin{align}
\frac{\mathrm{d}\sigma_{\mathrm{bg}}}
{\mathrm{d}y_\gamma\,\mathrm{d}^2k_\perp}
&=
\frac{\alpha_e^2}{2(2\pi)^3}
\int
\mathrm{d}y_1\,\mathrm{d}y_2\,
\mathrm{d}^2p_{1\perp}\,\mathrm{d}^2p_{2\perp}
\int
\mathrm{d}y_{\gamma^\ast}\,\mathrm{d}^2q_\perp\,
\frac{H_{\mathrm{Born}}}{Q^4}\,
f_{\mathrm{BH}}^{\gamma^\ast}
\left(y_{\gamma^\ast},Q^2,q_\perp^2\right)
\theta(-y_{\gamma^\ast})
\nonumber\\
&\qquad\times
\left[1+n(y_\gamma,|k_\perp|)\right]
\delta^{(4)}(q-p_1-p_2-k)
\nonumber\\
&=
\frac{\alpha_e^2}{2(2\pi)^3}
\int
\mathrm{d}y_1\,\mathrm{d}y_2\,
\mathrm{d}^2p_\perp\,\mathrm{d}^2P_\perp
\int
\mathrm{d}y_{\gamma^\ast}\,\mathrm{d}^2q_\perp\,
\frac{H_{\mathrm{Born}}}{Q^4}\,
f_{\mathrm{BH}}^{\gamma^\ast}
\left(y_{\gamma^\ast},Q^2,q_\perp^2\right)
\theta(-y_{\gamma^\ast})
\nonumber\\
&\qquad\times
\left[1+n(y_\gamma,|k_\perp|)\right]
\delta^{(4)}(q-p_1-p_2-k).
\label{eq:bg-photon-spectrum}
\end{align}
We similarly define the reference cross section in the absence of the photon background:
\begin{align}
\frac{\mathrm{d}\sigma_0}
{\mathrm{d}y_\gamma\,\mathrm{d}^2k_\perp}
&=
\frac{\alpha_e^2}{2(2\pi)^3}
\int
\mathrm{d}y_1\,\mathrm{d}y_2\,
\mathrm{d}^2p_{1\perp}\,\mathrm{d}^2p_{2\perp}
\int
\mathrm{d}y_{\gamma^\ast}\,\mathrm{d}^2q_\perp\,
\frac{H_{\mathrm{Born}}}{Q^4}\,
f_{\mathrm{BH}}^{\gamma^\ast}
\left(y_{\gamma^\ast},Q^2,q_\perp^2\right)
\nonumber\\
&\qquad\times\theta(-y_{\gamma^\ast})
\delta^{(4)}(q-p_1-p_2-k)
\nonumber\\
&=
\frac{\alpha_e^2}{2(2\pi)^3}
\int
\mathrm{d}y_1\,\mathrm{d}y_2\,
\mathrm{d}^2p_\perp\,\mathrm{d}^2P_\perp
\int
\mathrm{d}y_{\gamma^\ast}\,\mathrm{d}^2q_\perp\,
\frac{H_{\mathrm{Born}}}{Q^4}\,
f_{\mathrm{BH}}^{\gamma^\ast}
\left(y_{\gamma^\ast},Q^2,q_\perp^2\right)
\nonumber\\
&\qquad\times\theta(-y_{\gamma^\ast})
\delta^{(4)}(q-p_1-p_2-k).
\label{eq:vacuum-photon-spectrum}
\end{align}
We then define the ratio $R^f$ to extract the background photon distribution:
\begin{equation}
    R^f(y_\gamma,k_\perp)
    \equiv
    \frac{
    \mathrm{d}\sigma_{\mathrm{bg}}/
    (\mathrm{d}y_\gamma\,\mathrm{d}^2k_\perp)}
    {
    \mathrm{d}\sigma_0/
    (\mathrm{d}y_\gamma\,\mathrm{d}^2k_\perp)}
    =
    1+n(y_\gamma,|k_\perp|).
    \label{eq:ratio-extraction}
\end{equation}
The validity of Eq.~\eqref{eq:ratio-extraction} follows directly from the factorized structure of the background correction. Once $y_\gamma$ and $k_\perp$ are fixed, the factor $1+n(y_\gamma,|k_\perp|)$ is independent of all remaining integration variables and can therefore be taken outside the integrals. The remaining integrands in the numerator and denominator are identical and cancel in the ratio. The background photon distribution is thus given by
\begin{equation}
    n(y_\gamma,|k_\perp|)
    =
    R^f(y_\gamma,k_\perp)-1.
    \label{eq:photon-distribution-extraction}
\end{equation}
In kinematic regions with a large photon occupation number, $R^f\gg1$, and one has the approximate relation
\begin{equation}
    n(y_\gamma,|k_\perp|)
    \simeq
    R^f(y_\gamma,k_\perp).
    \label{eq:large-occupation-approximation}
\end{equation}

The numerator of $R^f$ is the differential cross section that can be measured experimentally, whereas the denominator is a theoretical reference cross section that contains no background-photon enhancement and can be calculated perturbatively. This procedure does not require an assumed parametrization of the background photon distribution and therefore reduces model dependence, making it suitable for complicated background fields. The ratio provides a direct mapping from the measured photon spectrum to the background photon distribution function. It is essential, however, that the numerator and denominator be evaluated with identical kinematic cuts and phase-space definitions; otherwise, Eq.~\eqref{eq:ratio-extraction} does not hold.

\section{Effects of the Sudakov Factor on the Extraction Method}
\label{sec4}
The physical process is more complicated than its Born-level description. After being produced, the electron--positron pair is not observed immediately but may undergo multiple scattering in the background field or radiate soft photons that escape detection. Here, we focus on the effect of unresolved soft-photon radiation, which is conventionally incorporated through a Sudakov factor in transverse-coordinate space. After including Sudakov resummation, the cross section becomes
\begin{align}
&\frac{\mathrm{d}\sigma_{\mathrm{bg}}^{\mathrm{Sud}}}
{\mathrm{d}y_1\,\mathrm{d}y_2\,\mathrm{d}y_\gamma\,
 \mathrm{d}^2p'_\perp\,\mathrm{d}^2P_\perp\,
 \mathrm{d}^2k_\perp}
\nonumber\\
&\quad =
\int\frac{\mathrm{d}^2r_\perp}{(2\pi)^2}\,
e^{i r_\perp\cdot p'_\perp}
e^{-S_{\mathrm{Sud}}(P_\perp,r_\perp)}
\int\mathrm{d}^2p_\perp\,
e^{-i r_\perp\cdot p_\perp}
\frac{\mathrm{d}\sigma_{\mathrm{bg}}}
{\mathrm{d}y_1\,\mathrm{d}y_2\,\mathrm{d}y_\gamma\,
 \mathrm{d}^2p_\perp\,\mathrm{d}^2P_\perp\,
 \mathrm{d}^2k_\perp},
\label{eq:sudakov-convolution}
\end{align}
where
\begin{equation}
    S_{\mathrm{Sud}}(P_\perp,r_\perp)
    =
    \frac{\alpha_e}{2\pi}
    \ln^2\left(\frac{P_\perp^2}{\mu_r^2}\right),
    \qquad
    \mu_r=\frac{2e^{-\gamma_E}}{|r_\perp|}.
    \label{eq:sudakov-factor}
\end{equation}
The differential cross section appearing on the right-hand side of Eq.~\eqref{eq:sudakov-convolution} is given by Eq.~\eqref{eq:bg-differential-cross-section} after transforming from the individual lepton transverse momenta to $p_\perp$ and $P_\perp$.

Using the transverse-momentum-conservation delta function, Eq.~\eqref{eq:sudakov-convolution} can be written as
\begin{align}
&\frac{\mathrm{d}\sigma_{\mathrm{bg}}^{\mathrm{Sud}}}
{\mathrm{d}y_1\,\mathrm{d}y_2\,\mathrm{d}y_\gamma\,
 \mathrm{d}^2p'_\perp\,\mathrm{d}^2P_\perp\,
 \mathrm{d}^2k_\perp}
\nonumber\\
&\quad =
\frac{\alpha_e^2}{2(2\pi)^3}
\int
\mathrm{d}y_{\gamma^\ast}\,\mathrm{d}^2q_\perp\,
\mathcal{K}_{\mathrm{Sud}}
\left(P_\perp,q_\perp-p'_\perp-k_\perp\right)
\frac{H_{\mathrm{Born}}}{Q^4}\,
f_{\mathrm{BH}}^{\gamma^\ast}
\left(y_{\gamma^\ast},Q^2,q_\perp^2\right)
\theta(-y_{\gamma^\ast})
\nonumber\\
&\qquad\times
\left[1+n(y_\gamma,|k_\perp|)\right]
\delta^{(+,-)}(q-p_1-p_2-k),
\label{eq:bg-sudakov-cross-section}
\end{align}
where $\delta^{(+,-)}$ denotes the delta functions enforcing momentum conservation in the two light-cone directions, and the transverse Sudakov kernel is defined by
\begin{equation}
    \mathcal{K}_{\mathrm{Sud}}(P_\perp,\Delta_\perp)
    \equiv
    \int\frac{\mathrm{d}^2r_\perp}{(2\pi)^2}\,
    e^{-S_{\mathrm{Sud}}(P_\perp,r_\perp)}
    e^{-i r_\perp\cdot\Delta_\perp}.
    \label{eq:sudakov-kernel}
\end{equation}
In Eq.~\eqref{eq:bg-sudakov-cross-section},
$\Delta_\perp=q_\perp-p'_\perp-k_\perp$.

When the final-state photon is measured at fixed $y_\gamma$ and $k_\perp$, the ratio including Sudakov effects can be defined as
\begin{equation}
    R_{\mathrm{Sud}}^f(y_\gamma,k_\perp)
    \equiv
    \frac{
    \mathrm{d}\sigma_{\mathrm{bg}}^{\mathrm{Sud}}/
    (\mathrm{d}y_\gamma\,\mathrm{d}^2k_\perp)}
    {
    \mathrm{d}\sigma_0^{\mathrm{Sud}}/
    (\mathrm{d}y_\gamma\,\mathrm{d}^2k_\perp)}
    =
    1+n(y_\gamma,|k_\perp|)
    =
    R^f(y_\gamma,k_\perp).
    \label{eq:sudakov-ratio}
\end{equation}
Thus, the background photon distribution can still be extracted from the cross-section ratio after Sudakov resummation. The appropriate theoretical reference is no longer the Born-level expression in Eq.~\eqref{eq:vacuum-photon-spectrum}, but the Sudakov-corrected vacuum cross section
\begin{align}
\frac{\mathrm{d}\sigma_0^{\mathrm{Sud}}}
{\mathrm{d}y_\gamma\,\mathrm{d}^2k_\perp}
&=
\frac{\alpha_e^2}{2(2\pi)^3}
\int
\mathrm{d}y_1\,\mathrm{d}y_2\,
\mathrm{d}^2p'_\perp\,\mathrm{d}^2P_\perp
\int
\mathrm{d}y_{\gamma^\ast}\,\mathrm{d}^2q_\perp\,
\mathcal{K}_{\mathrm{Sud}}
\left(P_\perp,q_\perp-p'_\perp-k_\perp\right)
\nonumber\\
&\qquad\times
\frac{H_{\mathrm{Born}}}{Q^4}\,
f_{\mathrm{BH}}^{\gamma^\ast}
\left(y_{\gamma^\ast},Q^2,q_\perp^2\right)
\theta(-y_{\gamma^\ast})
\delta^{(+,-)}(q-p_1-p_2-k).
\label{eq:vacuum-sudakov-spectrum}
\end{align}

The Sudakov factor itself therefore does not invalidate the cross-section-ratio inversion. What invalidates the inversion is an inconsistent theoretical treatment of the numerator and denominator. If the experimental numerator corresponds to the physical cross section including Sudakov corrections, while the theoretical denominator is evaluated only at Born level, the relation in Eq.~\eqref{eq:sudakov-ratio} is no longer valid. Schematically,
\begin{equation}
    \frac{\mathrm{d}\sigma_{\mathrm{bg}}^{\mathrm{Sud}}}
    {\mathrm{d}\sigma_0^{\mathrm{Born}}}
    \neq 1+n,
    \qquad \rm but \quad
    \frac{\mathrm{d}\sigma_{\mathrm{bg}}^{\mathrm{Sud}}}
    {\mathrm{d}\sigma_0^{\mathrm{Sud}}}
    =1+n.
    \label{eq:consistent-sudakov-treatment}
\end{equation}
The theoretical denominator must therefore be calculated at the same perturbative accuracy and with the same phase-space definition as the experimental numerator. The three relevant cases are summarized in Table~\ref{tab:ratio-inversion}.

\begin{table}[htbp]
    \centering
    \caption{Cross-section-ratio inversion under different theoretical treatments.}
    \label{tab:ratio-inversion}
    \begin{tabular}{cccc}
        \toprule
        Treatment & Numerator & Denominator & Result \\
        \midrule
        Born/Born
        & $\mathrm{d}\sigma_{\mathrm{bg}}$
        & $\mathrm{d}\sigma_0$
        & $R=1+n$ \\
        Sudakov/Sudakov
        & $\mathrm{d}\sigma_{\mathrm{bg}}^{\mathrm{Sud}}$
        & $\mathrm{d}\sigma_0^{\mathrm{Sud}}$
        & $R=1+n$ \\
        Sudakov/Born
        & $\mathrm{d}\sigma_{\mathrm{bg}}^{\mathrm{Sud}}$
        & $\mathrm{d}\sigma_0$
        & $R\neq1+n$ \\
        \bottomrule
    \end{tabular}
\end{table}

The role of the Sudakov factor is therefore not to invalidate the ratio method, but to require that the theoretical reference cross section without the background field be evaluated with the same radiative corrections as the measured cross section. Only under this consistency condition can the extracted distribution $n(y_\gamma,|k_\perp|)$ be guaranteed not to contain spurious structures induced by unresolved soft radiation.

\section{Effects of Kinematic Cuts, Finite Bin Widths, and Experimental Acceptance on the Extraction}
\label{sec5}
The preceding sections have shown that, at the level of ideal differential cross sections, the final-state real-photon propagator in the background field introduces only a multiplicative factor. Consequently, when the rapidity and transverse momentum of the final-state photon are fixed, the ratio of the cross section in the background field to the corresponding reference cross section without the background field directly determines the background photon occupation number. In this section, we discuss the conditions under which this relation remains applicable to realistic physical processes and experimental analyses. It should be emphasized that the cross-section-ratio inversion is not a fitting assumption, but an algebraic consequence of the multiplicative structure of the final-state photon propagator. The relevant experimental effects that must be controlled include kinematic cuts, finite bin widths, detector acceptance, and the treatment of unobserved degrees of freedom.

We first consider the fundamental requirement for the inversion relation, namely, the use of identical kinematic cuts. For compactness, we denote the measured final-state photon variables by
\begin{equation}
    X=(y_\gamma,\bm{k}_\perp).
    \label{eq:observable-X}
\end{equation}
In the presence of the background field, the leading-order cross section can be written schematically as
\begin{equation}
    \frac{\mathrm{d}\sigma_{\mathrm{bg}}}{\mathrm{d}X}
    =
    \int \mathrm{d}\Phi\,
    K(X,\Phi)\left[1+n(X)\right],
    \label{eq:bg-general-phase-space}
\end{equation}
where $\mathrm{d}\Phi$ denotes the phase-space variables other than $X$ that are integrated over, while $K(X,\Phi)$ contains the hard-scattering kernel, the virtual-photon source function, propagator factors, phase-space factors, and the momentum-conservation delta functions. The corresponding reference cross section without the background field is
\begin{equation}
    \frac{\mathrm{d}\sigma_0}{\mathrm{d}X}
    =
    \int \mathrm{d}\Phi\,K(X,\Phi).
    \label{eq:vacuum-general-phase-space}
\end{equation}
Because $n(X)$ depends only on the fixed final-state photon variables $X$ and not on the integrated variables $\Phi$, one obtains
\begin{equation}
    \frac{\mathrm{d}\sigma_{\mathrm{bg}}}{\mathrm{d}X}
    =
    \left[1+n(X)\right]
    \frac{\mathrm{d}\sigma_0}{\mathrm{d}X}.
    \label{eq:general-multiplicative-relation}
\end{equation}
The direct inversion relation is therefore
\begin{equation}
    n(X)
    =
    \frac{
    \mathrm{d}\sigma_{\mathrm{bg}}/\mathrm{d}X
    }{
    \mathrm{d}\sigma_0/\mathrm{d}X
    }
    -1.
    \label{eq:general-direct-inversion}
\end{equation}
For Eq.~\eqref{eq:general-direct-inversion} to hold, the factors $\int\mathrm{d}\Phi\,K(X,\Phi)$ in the numerator and denominator must cancel exactly. This cancellation requires the integration regions in the remaining phase space $\mathrm{d}\Phi$ to be identical. The use of identical phase-space definitions is therefore a fundamental condition of the method. For example, if events in a particular kinematic region are lost or removed from the experimental sample, the same region must also be excluded from the $\mathrm{d}\Phi$ integration used to construct the denominator of the ratio.

We next consider the effect of finite bin widths. A real experiment cannot measure an infinitesimal differential cross section, but only an integrated cross section over a finite bin. For a bin $\Delta X$, the measured background-field cross section is
\begin{equation}
    \sigma_{\mathrm{bg}}^{\Delta X}
    =
    \int_{\Delta X}\mathrm{d}X
    \int\mathrm{d}\Phi\,
    K(X,\Phi)\left[1+n(X)\right].
    \label{eq:bg-finite-bin}
\end{equation}
The corresponding bin-averaged cross section is
\begin{equation}
    \overline{\sigma}_{\mathrm{bg}}^{\Delta X}
    =
    \frac{
    \displaystyle
    \int_{\Delta X}\mathrm{d}X
    \int\mathrm{d}\Phi\,
    K(X,\Phi)\left[1+n(X)\right]
    }{
    \displaystyle
    \int_{\Delta X}\mathrm{d}X
    }.
    \label{eq:bg-bin-average}
\end{equation}
In the infinitesimal-bin limit,
\begin{equation}
    \lim_{\Delta X\rightarrow 0}
    \overline{\sigma}_{\mathrm{bg}}^{\Delta X}
    =
    \left.
    \frac{\mathrm{d}\sigma_{\mathrm{bg}}}{\mathrm{d}X}
    \right|_{X=X_0},
    \label{eq:bg-bin-limit}
\end{equation}
where $X_0$ denotes the central value of the bin. Thus, when $\Delta X$ is sufficiently small, the experimentally measured bin average approaches the differential cross section required in the numerator of the ratio.

The reference cross section without the background field is defined analogously as
\begin{equation}
    \sigma_0^{\Delta X}
    =
    \int_{\Delta X}\mathrm{d}X
    \int\mathrm{d}\Phi\,K(X,\Phi),
    \label{eq:vacuum-finite-bin}
\end{equation}
with the bin-averaged value
\begin{equation}
    \overline{\sigma}_0^{\Delta X}
    =
    \frac{
    \displaystyle
    \int_{\Delta X}\mathrm{d}X
    \int\mathrm{d}\Phi\,K(X,\Phi)
    }{
    \displaystyle
    \int_{\Delta X}\mathrm{d}X
    }.
    \label{eq:vacuum-bin-average}
\end{equation}
The ratio evaluated within the finite bin is then
\begin{align}
    R_f^{\Delta X}
    &\equiv
    \frac{
    \overline{\sigma}_{\mathrm{bg}}^{\Delta X}
    }{
    \overline{\sigma}_0^{\Delta X}
    }
    =
    \frac{
    \sigma_{\mathrm{bg}}^{\Delta X}
    }{
    \sigma_0^{\Delta X}
    }
   =
    1+
    \frac{
    \displaystyle
    \int_{\Delta X}\mathrm{d}X
    \int\mathrm{d}\Phi\,
    K(X,\Phi)n(X)
    }{
    \displaystyle
    \int_{\Delta X}\mathrm{d}X
    \int\mathrm{d}\Phi\,
    K(X,\Phi)
    }.
    \label{eq:finite-bin-ratio}
\end{align}
Equation~\eqref{eq:finite-bin-ratio} shows that a finite-bin measurement does not extract the pointwise value $n(X)$, but rather a weighted average of $n(X)$ over the region $\Delta X$, with the reference cross section determining the weight. Defining
\begin{equation}
    n^{\Delta X}(X)
    \equiv
    R_f^{\Delta X}(X)-1,
    \label{eq:finite-bin-distribution}
\end{equation}
one finds
\begin{equation}
    \lim_{\Delta X\rightarrow 0}
    n^{\Delta X}(X)
    =
    n(X).
    \label{eq:finite-bin-distribution-limit}
\end{equation}
Therefore, if the bin is sufficiently narrow and both $n(X)$ and the hard-process weight $\int\mathrm{d}\Phi\,K(X,\Phi)$ vary slowly within the bin, one may approximate
\begin{equation}
    n(X)\simeq n^{\Delta X}(X),
    \label{eq:narrow-bin-approximation}
\end{equation}
where $X$ is taken to be the bin center. A finite bin width therefore does not invalidate the ratio method, but converts a pointwise inversion into a cross-section-weighted average. In an experimental analysis, sufficiently narrow bins in $y_\gamma$ and $|\bm{k}_\perp|$ should be used whenever statistically feasible, and the theoretical denominator must in all cases be integrated over exactly the same bins as the experimental numerator. For a broad bin extending from $X$ to $X+\Delta X$, the extracted quantity should be interpreted as the average of $n(X)$ over that interval, weighted by the hard-process kernel $\int\mathrm{d}\Phi\,K(X,\Phi)$.

Finally, we consider experimental acceptance. Detector acceptance and analysis cuts modify the weighting of the phase-space integral. Let $A(X,\Phi)$ denote the detector acceptance and selection-efficiency function. The experimentally observable background-field cross section is then
\begin{equation}
    \frac{\mathrm{d}\sigma_{\mathrm{bg}}^{\mathrm{acc}}}
    {\mathrm{d}X}
    =
    \int\mathrm{d}\Phi\,
    A(X,\Phi)K(X,\Phi)
    \left[1+n(X)\right].
    \label{eq:bg-acceptance-cross-section}
\end{equation}
The corresponding theoretical reference cross section must be defined as
\begin{equation}
    \frac{\mathrm{d}\sigma_0^{\mathrm{acc}}}
    {\mathrm{d}X}
    =
    \int\mathrm{d}\Phi\,
    A(X,\Phi)K(X,\Phi).
    \label{eq:vacuum-acceptance-cross-section}
\end{equation}
Provided that exactly the same acceptance function is applied to the numerator and denominator, one still obtains
\begin{equation}
    R_{\mathrm{acc}}(X)
    \equiv
    \frac{
    \mathrm{d}\sigma_{\mathrm{bg}}^{\mathrm{acc}}/\mathrm{d}X
    }{
    \mathrm{d}\sigma_0^{\mathrm{acc}}/\mathrm{d}X
    }
    =
    1+n(X).
    \label{eq:acceptance-ratio}
\end{equation}
Experimental acceptance is therefore not intrinsically problematic. The relevant issue is whether the numerator and denominator are treated consistently. If a cut is applied to the experimental numerator but omitted from the theoretical denominator, the ratio acquires an additional phase-space weighting and may develop artificial structures. In practical applications, the theoretical denominator should not be interpreted as a simple inclusive Born-level cross section, but as the reference cross section without the background field evaluated under precisely the same experimental definition, including identical cuts, acceptance, efficiencies, and binning. Under these conditions,
\begin{equation}
    n(X)
    =
    R_{\mathrm{acc}}(X)-1
    =
    \frac{
    \mathrm{d}\sigma_{\mathrm{bg}}^{\mathrm{acc}}/\mathrm{d}X
    }{
    \mathrm{d}\sigma_0^{\mathrm{acc}}/\mathrm{d}X
    }
    -1.
    \label{eq:acceptance-extraction}
\end{equation}

\section{Summary and Outlook}
\label{sec6}
In this work, we have proposed a distribution-function inversion method based on ratios of differential cross sections, with the aim of reducing the dependence of conventional global analyses on predetermined functional parametrizations. As a proof of concept, we considered a QED analogue in electron--ion collisions. A Bethe--Heitler-type subprocess first produces a timelike virtual photon, which decays into a lepton pair. The lepton pair subsequently radiates a real photon while propagating through the coherent electromagnetic field of the nucleus. By analyzing the factorized structure of the complete cross section, we found that, once the rapidity and transverse momentum of the final-state photon are fixed, the original convolution can be reduced to a multiplicative form under the relevant approximations. This makes it possible to define a ratio between the experimentally measured cross section and a theoretically calculable hard reference,$R=\frac{\mathrm{d}\sigma_{\mathrm{exp}}}{\mathrm{d}\sigma_{\mathrm{hard}}},$ and to establish a direct correspondence between this ratio and the background photon distribution. Here, $\mathrm{d}\sigma_{\mathrm{hard}}$ includes the Bethe--Heitler virtual-photon source, the virtual-photon decay, phase-space factors, and all other perturbatively calculable contributions. The method permits a point-by-point reconstruction of the distribution function at different kinematic coordinates without requiring an assumed global functional form. It may therefore provide a complementary and less model-dependent local constraint to conventional parametrized fits.

Under realistic experimental conditions, the inversion relation is affected by soft-photon radiation, finite bin widths, detector acceptance, finite resolution, and integrations over unobserved degrees of freedom. Consequently, the theoretical input $\mathrm{d}\sigma_{\mathrm{hard}}$ cannot be identified merely with a lowest-order partonic cross section. It must instead be constructed using the same kinematic cuts, bin integrations, radiative corrections, and detector response as those applied to the experimental data. Future work will perform closure tests using Monte Carlo pseudodata and systematically investigate the effects of statistical uncertainties, finite detector resolution, higher-order QED corrections, nuclear form factors, and competing reaction channels on the stability of the inversion. The method will also be extended to QCD processes, such as radiation from a quark or diquark system propagating through the nuclear gluon background. If the corresponding factorization structure can be established, cross-section-ratio inversion may provide a new methodological tool for extracting nucleon and nuclear structure distributions at the EIC.

\section*{Acknowledgments}

\appendix

\section{Photon Propagator Modified by a Background Field}
\label{app:bg-photon-propagator}

In vacuum quantum field theory, the fundamental propagating object of the photon is the time-ordered two-point function
$\langle 0|T A_\mu(x)A_\nu(y)|0\rangle$.
However, in a strong background-field environment, scattering no longer takes place on the vacuum ground state, but in a statistical background state with a finite occupation number. It is therefore necessary to reconsider the structure of the gauge-field two-point function. For example, in heavy-ion collisions, $B$ corresponds to the equivalent photon field surrounding the heavy ion. In this case, the photon propagator can be decomposed into two parts:
\begin{equation}
D_{\mu\nu}^{\mathrm{bg}}
=
\langle 0|T A_\mu(x)A_\nu(y)|0\rangle
+
\langle B|T A_\mu(x)A_\nu(y)|B\rangle .
\end{equation}
Here, the first term is the familiar vacuum propagator, whereas the second term originates from the contribution of the background real-photon field. Furthermore, we have
\begin{align}
\langle B|T A_\mu(x)A_\nu(y)|B\rangle
&=
\operatorname{Tr}
\left[
\rho\,T A_\mu(x)A_\nu(y)
\right]
\notag\\
&=
\sum_{\lambda}
\int
\frac{\mathrm{d}^3 k}{(2\pi)^3 2k_0}\,
n(\bm{k})\,
\operatorname{Tr}
\left[
|k\rangle\langle k|
T A_\mu(x)A_\nu(y)
\right]\label{bgpro}\\
&=
\sum_{\lambda}
\int
\frac{\mathrm{d}^3 k}{(2\pi)^3 2k_0}\,
n(\bm{k})
\langle k|
T A_\mu(x)A_\nu(y)
|k\rangle
\notag\\
&=
\int
\frac{\mathrm{d}^3 k}{(2\pi)^3 2k_0}\,
n(\bm{k})e^{-ik\cdot(x-y)}
\theta(x^0-y^0)
\sum_{\lambda}
\epsilon_\mu^{(\lambda)*}
\epsilon_\nu^\lambda
+
(x\leftrightarrow y)
\notag\\
&=
\int
\frac{\mathrm{d}^3 k}{(2\pi)^3 2k_0}\,
n(\bm{k})e^{-ik\cdot(x-y)}
\theta(x^0-y^0)
(-g_{\mu\nu})
+
(x\leftrightarrow y)
\notag\\
&=
\int
\frac{\mathrm{d}^4 k}{(2\pi)^4}\,
e^{-ik\cdot(x-y)}
\frac{-ig_{\mu\nu}}{k^2+i\varepsilon}
n(\bm{k}).
\notag
\end{align}
Here,
\begin{equation}
A_\nu(y)|k\rangle
=
\epsilon_\nu^\lambda(k)e^{-ik\cdot y}|0\rangle,
\end{equation}
and
\begin{equation}
\rho
=
|B\rangle\langle B|
=
\sum_{\lambda}
\int
\frac{\mathrm{d}^3 k}{(2\pi)^3 2k_0}\,
n(\bm{k})|k\rangle\langle k|
\end{equation}
is the density matrix of the statistical background state. As can be seen from Eq.~\eqref{bgpro}, the contribution of the statistical background state is equivalent to introducing, into the vacuum-propagator structure, a statistical weight controlled by the occupation-number function $n(\bm{k})$. This allows the background effect to enter the amplitude calculation through the two-point function, without introducing empirical parameters at the level of the final result. In the second equality of Eq.~\eqref{bgpro}, we have used the standard residue theorem to combine the two time-ordered terms into $1/(k^2+i\epsilon)$. Here, $n(\bm{k})$ is assumed to be an even function. Physically, $n(\bm{k})$ represents the mean photon occupation number and corresponds to the number of real photons with momentum magnitude $|k|=E_k$. It can be clearly seen that, in a real-photon background field, the photon propagator is modified relative to its vacuum form. It is equivalent to a convolution of the vacuum propagator with the occupation-number function $n(\bm{k})$, indicating that the photon-exchange process is affected by the absolute number of real photons. Therefore, the photon propagator in the background field can be written as
\begin{equation}
D_{\mu\nu}^{\mathrm{bg}}(x-y)
=
\int
\frac{\mathrm{d}^4 k}{(2\pi)^4}\,
e^{-ik\cdot(x-y)}
\frac{-ig_{\mu\nu}}{k^2+i\varepsilon}
\left[n(\bm{k})+1\right].
\end{equation}
The corresponding on-shell/cut propagator in momentum-space is
\begin{equation}
    D_{\mu\nu}^{\mathrm{cut};\mathrm{bg}}(k)
    =
    2\pi\theta(k^0)\delta(k^2)
    \left(-g_{\mu\nu}\right)
    \left[1+n(\bm{k})\right],
    \label{eq:appendix-cut-bg-propagator}
\end{equation}
which is the expression used in the main text.

\bibliography{ref}

@article{Collins:1985,
title = {Factorization for short distance hadron-hadron scattering},
journal = {Nuclear Physics B},
volume = {261},
pages = {104-142},
year = {1985},
issn = {0550-3213},
doi = {https://doi.org/10.1016/0550-3213(85)90565-6},
url = {https://www.sciencedirect.com/science/article/pii/0550321385905656},
author = {John C. Collins and Davison E. Soper and George Sterman}
}

@article{Gao:2018,
title = {The structure of the proton in the LHC precision era},
journal = {Physics Reports},
volume = {742},
pages = {1-121},
year = {2018},
note = {The Structure of the Proton in the LHC Precision Era},
issn = {0370-1573},
doi = {https://doi.org/10.1016/j.physrep.2018.03.002},
url = {https://www.sciencedirect.com/science/article/pii/S0370157318300371},
author = {Jun Gao and Lucian Harland-Lang and Juan Rojo},
}

@article{DelDebbio:2022,
  author   = {Del Debbio, Luigi and Giani, Tommaso and Wilson, Michael},
  title    = {Bayesian approach to inverse problems: an application to NNPDF closure testing},
  journal  = {The European Physical Journal C},
  year     = {2022},
  volume   = {82},
  number   = {4},
  pages    = {330},
  doi      = {10.1140/epjc/s10052-022-10297-x},
  url      = {https://doi.org/10.1140/epjc/s10052-022-10297-x},
  issn     = {1434-6052},
}

@article{Ball:2022,
  author  = {Ball, Richard D. and Carrazza, Stefano and Cruz-Martinez, Juan and Del Debbio, Luigi and Forte, Stefano and Giani, Tommaso and Iranipour, Shayan and Kassabov, Zahari and Latorre, Jose I. and Nocera, Emanuele R. and Pearson, Rosalyn L. and Rojo, Juan and Stegeman, Roy and Schwan, Christopher and Ubiali, Maria and Voisey, Cameron and Wilson, Michael},
  title   = {The path to proton structure at 1\% accuracy},
  journal = {The European Physical Journal C},
  year    = {2022},
  volume  = {82},
  number  = {5},
  pages   = {428},
  doi     = {10.1140/epjc/s10052-022-10328-7},
  url     = {https://doi.org/10.1140/epjc/s10052-022-10328-7},
  issn    = {1434-6052}
}

@article{Eskola:2022,
  author   = {Eskola, Kari J. and Paakkinen, Petja and Paukkunen, Hannu and Salgado, Carlos A.},
  title    = {EPPS21: a global QCD analysis of nuclear PDFs},
  journal  = {The European Physical Journal C},
  year     = {2022},
  volume   = {82},
  number   = {5},
  pages    = {413},
  doi      = {10.1140/epjc/s10052-022-10359-0},
  url      = {https://doi.org/10.1140/epjc/s10052-022-10359-0},
  issn     = {1434-6052},
}

@article{MAP:2022,
  author  = {Bacchetta, Alessandro and Bertone, Valerio and Bissolotti, Chiara and Bozzi, Giuseppe and Cerutti, Matteo and Piacenza, Fulvio and Radici, Marco and Signori, Andrea and {The MAP Collaboration}},
  title   = {Unpolarized transverse momentum distributions from a global fit of Drell-Yan and semi-inclusive deep-inelastic scattering data},
  journal = {Journal of High Energy Physics},
  year    = {2022},
  volume  = {2022},
  number  = {10},
  pages   = {127},
  doi     = {10.1007/JHEP10(2022)127},
  url     = {https://doi.org/10.1007/JHEP10(2022)127},
  issn    = {1029-8479}
}

@article{deFlorian:2009,
  title = {Extraction of spin-dependent parton densities and their uncertainties},
  author = {de Florian, Daniel and Sassot, Rodolfo and Stratmann, Marco and Vogelsang, Werner},
  journal = {Phys. Rev. D},
  volume = {80},
  issue = {3},
  pages = {034030},
  numpages = {26},
  year = {2009},
  month = {Aug},
  publisher = {American Physical Society},
  doi = {10.1103/PhysRevD.80.034030},
  url = {https://link.aps.org/doi/10.1103/PhysRevD.80.034030}
}

@article{Ball:2015,
  author  = {Ball, Richard D. and Bertone, Valerio and Carrazza, Stefano and Deans, Christopher S. and Del Debbio, Luigi and Forte, Stefano and Guffanti, Alberto and Hartland, Nathan P. and Latorre, Jos{\'e} I. and Rojo, Juan and Ubiali, Maria and {The NNPDF collaboration}},
  title   = {Parton distributions for the LHC run II},
  journal = {Journal of High Energy Physics},
  year    = {2015},
  volume  = {2015},
  number  = {4},
  pages   = {40},
  doi     = {10.1007/JHEP04(2015)040},
  url     = {https://doi.org/10.1007/JHEP04(2015)040},
  issn    = {1029-8479}
}

@article{Carrazza2022,
  author  = {Carrazza, Stefano and Cruz-Martinez, Juan and Stegeman, Roy},
  title   = {A data-based parametrization of parton distribution functions},
  journal = {The European Physical Journal C},
  year    = {2022},
  volume  = {82},
  number  = {2},
  pages   = {163},
  doi     = {10.1140/epjc/s10052-022-10136-z},
  url     = {https://doi.org/10.1140/epjc/s10052-022-10136-z},
  issn    = {1434-6052}
}

@article{Khalek2019,
  author  = {Khalek, Rabah Abdul and Ethier, Jacob J. and Rojo, Juan},
  title   = {Nuclear parton distributions from lepton-nucleus scattering and the impact of an electron-ion collider},
  journal = {The European Physical Journal C},
  year    = {2019},
  volume  = {79},
  number  = {6},
  pages   = {471},
  doi     = {10.1140/epjc/s10052-019-6983-1},
  url     = {https://doi.org/10.1140/epjc/s10052-019-6983-1},
  issn    = {1434-6052}
}

@article{Scimemi2020,
  author  = {Scimemi, Ignazio and Vladimirov, Alexey},
  title   = {Non-perturbative structure of semi-inclusive deep-inelastic and Drell-Yan scattering at small transverse momentum},
  journal = {Journal of High Energy Physics},
  year    = {2020},
  volume  = {2020},
  number  = {6},
  pages   = {137},
  doi     = {10.1007/JHEP06(2020)137},
  url     = {https://doi.org/10.1007/JHEP06(2020)137},
  issn    = {1029-8479}
}

@article{Bury2022,
  author  = {Bury, Marcin and Hautmann, Francesco and Leal-Gomez, Sergio and Scimemi, Ignazio and Vladimirov, Alexey and Zurita, Pia},
  title   = {PDF bias and flavor dependence in TMD distributions},
  journal = {Journal of High Energy Physics},
  year    = {2022},
  volume  = {2022},
  number  = {10},
  pages   = {118},
  doi     = {10.1007/JHEP10(2022)118},
  url     = {https://doi.org/10.1007/JHEP10(2022)118},
  issn    = {1029-8479}
}

@article{PhysRevD.83.105005,
  title = {Universality of unintegrated gluon distributions at small $x$},
  author = {Dominguez, Fabio and Marquet, Cyrille and Xiao, Bo-Wen and Yuan, Feng},
  journal = {Phys. Rev. D},
  volume = {83},
  issue = {10},
  pages = {105005},
  numpages = {27},
  year = {2011},
  month = {May},
  publisher = {American Physical Society},
  doi = {10.1103/PhysRevD.83.105005},
  url = {https://link.aps.org/doi/10.1103/PhysRevD.83.105005}
}

@article{PhysRevLett.132.081902,
  title = {Back-to-Back Inclusive Dijets in Deep Inelastic Scattering at Small $x$: Complete NLO Results and Predictions},
  author = {Caucal, Paul and Salazar, Farid and Schenke, Bj\"orn and Stebel, Tomasz and Venugopalan, Raju},
  journal = {Phys. Rev. Lett.},
  volume = {132},
  issue = {8},
  pages = {081902},
  numpages = {8},
  year = {2024},
  month = {Feb},
  publisher = {American Physical Society},
  doi = {10.1103/PhysRevLett.132.081902},
  url = {https://link.aps.org/doi/10.1103/PhysRevLett.132.081902}
}

@article{Sun_2020,
   title={Studying Coulomb correction at EIC and EicC},
   volume={808},
   ISSN={0370-2693},
   url={http://dx.doi.org/10.1016/j.physletb.2020.135679},
   DOI={10.1016/j.physletb.2020.135679},
   journal={Physics Letters B},
   publisher={Elsevier BV},
   author={Sun, Ze-hao and Zheng, Du-xin and Zhou, Jian and Zhou, Ya-jin},
   year={2020},
   month=Sept, pages={135679} }

@article{ABDU,
  author  = {R. {Abdul Khalek} and others},
  title   = {Science Requirements and Detector Concepts for the Electron-Ion Collider: EIC Yellow Report},
  journal = {Nuclear Physics A},
  volume  = {1026},
  pages   = {122447},
  year    = {2022},
  issn    = {0375-9474},
  doi     = {10.1016/j.nuclphysa.2022.122447},
  url     = {https://www.sciencedirect.com/science/article/pii/S0375947422000677}
}

@article{Accardi2016,
  author  = {Accardi, A. and others},
  title   = {Electron-Ion Collider: The next QCD frontier},
  journal = {The European Physical Journal A},
  year    = {2016},
  volume  = {52},
  number  = {9},
  pages   = {268},
  doi     = {10.1140/epja/i2016-16268-9},
  url     = {https://doi.org/10.1140/epja/i2016-16268-9},
  issn    = {1434-601X}
}

@article{BUDNEV1975181,
title = {The two-photon particle production mechanism. Physical problems. Applications. Equivalent photon approximation},
journal = {Physics Reports},
volume = {15},
number = {4},
pages = {181-282},
year = {1975},
issn = {0370-1573},
doi = {https://doi.org/10.1016/0370-1573(75)90009-5},
url = {https://www.sciencedirect.com/science/article/pii/0370157375900095},
author = {V.M. Budnev and I.F. Ginzburg and G.V. Meledin and V.G. Serbo},
}

@article{BERTULANI1988299,
title = {Electromagnetic processes in relativistic heavy ion collisions},
journal = {Physics Reports},
volume = {163},
number = {5},
pages = {299-408},
year = {1988},
issn = {0370-1573},
doi = {https://doi.org/10.1016/0370-1573(88)90142-1},
url = {https://www.sciencedirect.com/science/article/pii/0370157388901421},
author = {Carlos A. Bertulani and Gerhard Baur},
}

@article{Jentschura2009,
  author  = {Jentschura, U. D. and Serbo, V. G.},
  title   = {Nuclear form factor, validity of the equivalent photon approximation and Coulomb corrections to muon pair production in photon--nucleus and nucleus--nucleus collisions},
  journal = {The European Physical Journal C},
  year    = {2009},
  volume  = {64},
  number  = {2},
  pages   = {309},
  doi     = {10.1140/epjc/s10052-009-1147-3},
  url     = {https://doi.org/10.1140/epjc/s10052-009-1147-3},
  issn    = {1434-6052}
}

@article{PhysRevD.101.075049,
  title = {Dark photon manifestation in the tripletlike QED processes $\ensuremath{\gamma}+{\ensuremath{\ell}}_{i}\ensuremath{\rightarrow}{\ensuremath{\ell}}_{j}^{+}{\ensuremath{\ell}}_{j}^{\ensuremath{-}}+{\ensuremath{\ell}}_{i}$, $i\ensuremath{\ne}j$, $i=e$, $\ensuremath{\mu}$, $j=e$, $\ensuremath{\mu}$, $\ensuremath{\tau}$},
  author = {Gakh, G. I. and Konchatnij, M. I. and Merenkov, N. P. and Tomasi--Gustafsson, E.},
  journal = {Phys. Rev. D},
  volume = {101},
  issue = {7},
  pages = {075049},
  numpages = {16},
  year = {2020},
  month = {Apr},
  publisher = {American Physical Society},
  doi = {10.1103/PhysRevD.101.075049},
  url = {https://link.aps.org/doi/10.1103/PhysRevD.101.075049}
}

@article{YENNIE1961379,
title = {The infrared divergence phenomena and high-energy processes},
journal = {Annals of Physics},
volume = {13},
number = {3},
pages = {379-452},
year = {1961},
issn = {0003-4916},
doi = {https://doi.org/10.1016/0003-4916(61)90151-8},
url = {https://www.sciencedirect.com/science/article/pii/0003491661901518},
author = {D.R Yennie and S.C Frautschi and H Suura},
}

@article{PhysRevD.102.094013,
  title = {Lepton pair production through two photon process in heavy ion collisions},
  author = {Klein, Spencer and Mueller, A. H. and Xiao, Bo-Wen and Yuan, Feng},
  journal = {Phys. Rev. D},
  volume = {102},
  issue = {9},
  pages = {094013},
  numpages = {22},
  year = {2020},
  month = {Nov},
  publisher = {American Physical Society},
  doi = {10.1103/PhysRevD.102.094013},
  url = {https://link.aps.org/doi/10.1103/PhysRevD.102.094013}
}

@article{PhysRevD.108.116015,
  title = {Lepton pair production in ultraperipheral collisions: Toward a precision test of the resummation formalism},
  author = {Shao, Ding Yu and Zhang, Cheng and Zhou, Jian and Zhou, Ya-jin},
  journal = {Phys. Rev. D},
  volume = {108},
  issue = {11},
  pages = {116015},
  numpages = {13},
  year = {2023},
  month = {Dec},
  publisher = {American Physical Society},
  doi = {10.1103/PhysRevD.108.116015},
  url = {https://link.aps.org/doi/10.1103/PhysRevD.108.116015}
}

@article{Li1,
   title={Probing the linear polarization of photons in ultraperipheral heavy ion collisions},
   volume={795},
   ISSN={0370-2693},
   url={http://dx.doi.org/10.1016/j.physletb.2019.07.005},
   DOI={10.1016/j.physletb.2019.07.005},
   journal={Physics Letters B},
   publisher={Elsevier BV},
   author={Li, Cong and Zhou, Jian and Zhou, Ya-jin},
   year={2019},
   month=aug, pages={576–580} }

@article{Li2,
   title={Impact parameter dependence of the azimuthal asymmetry in lepton pair production in heavy ion collisions},
   volume={101},
   ISSN={2470-0029},
   url={http://dx.doi.org/10.1103/PhysRevD.101.034015},
   DOI={10.1103/physrevd.101.034015},
   number={3},
   journal={Physical Review D},
   publisher={American Physical Society (APS)},
   author={Li, Cong and Zhou, Jian and Zhou, Ya-jin},
   year={2020},
   month=feb }
\end{document}